 \documentclass[final,5p,times,twocolumn]{elsarticle}
\pdfoutput=1
\usepackage{amsmath,amssymb,amsfonts,booktabs}
\usepackage{slashed}
\usepackage{bbm}
\usepackage{nicefrac}
\usepackage{dsfont}
\usepackage{fixmath}
\usepackage{mathtools}
\usepackage{picture}
\usepackage{amssymb}

\usepackage{microtype}
\usepackage{hyperref}
\usepackage{tikz}
\usetikzlibrary{shapes,arrows,automata,positioning}
\usepackage{tikz-3dplot}

\newcommand{\alg}[1]{\mathfrak{#1}}
\newcommand{\grp}[1]{\mathrm{#1}}

\DeclareMathOperator{\idm}{\mathds{1}}
\newcommand{\phicl}{\ensuremath{\phi}^{\mathrm{cl}}}
\newcommand{\varphicl}{\ensuremath{\varphi}^{\mathrm{cl}}}
\DeclareMathOperator{\pexp}{\mathrm{Pexp}}
\newcommand{\Ucl}{\ensuremath{U^{\mathrm{cl}}}}
\newcommand{\calA}{\ensuremath{\mathcal{A}}}
\DeclareMathOperator{\tr}{\mathrm{tr}}

\journal{Physics Letters B}
\begin{document}

\begin{frontmatter}



\title{Wilson lines in AdS/dCFT}

\author{Sara Bonansea, Khalil Idiab, Charlotte Kristjansen, and Matthias Volk}

\address[label2]{Niels Bohr Institute, Copenhagen University,\\ Blegdamsvej 17, 2100 Copenhagen \O, Denmark}
 


\begin{abstract}
We consider the expectation value of Wilson lines in two defect versions of ${\cal N}=4$ SYM, 
both with supersymmetry completely broken, where one is described in terms of an integrable boundary state, the other one not. For both cases, imposing a certain double scaling limit,  we find agreement to two leading orders between the expectation values calculated from respectively the field theory- and the string theory side of the AdS/dCFT correspondence.
 \end{abstract}


\end{frontmatter}



\section{Introduction}
\label{sec:introduction}

Understanding the interplay between supersymmetry and integrability in the AdS/CFT correspondence might hold the key
to understanding the deeper reason for the integrability of the systems involved.  Motivated by such considerations we will be pursuing
a line of investigation which involves breaking the supersymmetry of ${\cal N}=4$ SYM
in a simple way by introducing  a domain wall, a co-dimension one defect, separating two regions of space-time with different vacuum expectation values (vevs) for the scalar fields.  To be more precise, we will assign vevs in a particular way to either five
or to all six of the scalar fields on one side of the defect while keeping the vevs zero on the other side. 
In the language of integrability the defect can be described as a matrix product state or
a boundary state~\cite{deLeeuw:2015hxa} and for one of the set-ups the boundary state has been found to be integrable~\cite{deLeeuw:2018mkd}, for the other
one not~\cite{deLeeuw:2019sew}, where the notion of integrability of a matrix product state 
was introduced in~\cite{Piroli:2017sei}.
 The string theory duals of these defect CFTs are two D3-D7 probe brane systems, named I and II, with non-vanishing background gauge field flux and instanton number respectively, cf.\ table~\ref{Configurations}.

 Our aim will be to calculate a non-local observable,  the expectation value of a Wilson line, running parallel to the defect, both
from the gauge theory- and the string theory perspective. A double scaling limit, invented for a related supersymmetric D3-D5 probe
brane set-up in~\cite{Nagasaki:2012re} and generalized to the the two relevant D3-D7 probe brane set-ups in~\cite{Kristjansen:2012tn} will allow us to compare
the results of the two calculations. We remark that the gauge theory calculalations are rather involved due to the non-vanishing vevs which
mix color as well as flavor components of the ${\cal N}=4$ SYM fields but the perturbative framework necessary for the calculations has been set up in~\cite{Grau:2018keb} and~\cite{Gimenez-Grau:2019fld}. 

Earlier studies of Wilson loops in domain wall versions of ${\cal N}=4$ SYM have been limited to the supersymmetric and
integrable case of the D3-D5 probe brane system. For the D3-D5 case using the perturbative set-up developed in~\cite{Buhl-Mortensen:2016pxs,Buhl-Mortensen:2016jqo}, agreement between gauge and string theory calculations in
the double scaling limit was found for a single Wilson line in~\cite{Nagasaki:2011ue,deLeeuw:2016vgp}, a pair of Wilson lines in ~\cite{Preti:2017fhw} and a circular Wilson loop in~\cite{Bonansea:2019rxh}, see also~\cite{Aguilera-Damia:2016bqv}. 

With the present work we are able to address AdS/dCFT while eliminating both supersymmetry and (boundary) integrability. Interestingly, we find agreement between the gauge- and string theory result to two leading orders in the double scaling parameter for both of the
non-supersymmetric set-ups and in particular both for the integrable and the non-integrable case.

Our paper is organized in the following simple way. In section~\ref{sec:gauge-theory-computation} we compute the expectation value of the Wilson
line for our two defect set-ups from the gauge theory perspective whereafter in section~\ref{sec:string-theory-computation} we perform the computations
from the string theory perspective.  Finally, section~\ref{sec:conclusion} contains our conclusion and outlook.
{%
\renewcommand*{\arraystretch}{1.5}
\renewcommand{\tabcolsep}{1em}
\begin{table}
\begin{center}
\label{tab:results}
\begin{tabular}{l  c c}
    \toprule
D3-D7 set-up & I & II\\ 
     \midrule 
Supersymmetry & None & None  \\
Brane geometry  & AdS$_4\times $ S$^2$ $\times$ S$^2$& AdS$_4\times $ S$^4$  \\
Flux/Instanton no. & $k_1, k_2$ & $\frac{(n+1)(n+2)(n+3)} {6}$ \\
D.s. parameter    & $\frac{\lambda}{\pi^2(k_1^2+k_2^2)}$& $\frac{\lambda}{\pi^2n^2}$  \\
Boundary state  & Non-integrable & Integrable \\
\bottomrule 
\end{tabular}
\end{center}
\caption{\label{Configurations}The probe brane configurations dual to the dCFT versions of ${\cal N}=4$ SYM theory 
considered in this paper and their corresponding double scaling (d.s.) parameters. The
discussion of the integrability properties of the associated boundary states can be found in~\cite{deLeeuw:2018mkd,deLeeuw:2019sew}. }
\end{table}
}

\section{The Gauge Theory Computation}
\label{sec:gauge-theory-computation}

\subsection{The defect theories}
\label{eq:defect-theories}

The gauge theory duals of the two probe-brane setups of table~\ref{Configurations} are obtained as defect versions of $\mathcal{N} = 4$ SYM in which (some of) the scalar fields are assigned a non-vanishing vacuum expectation value for $x_3>0$.
The vevs are solutions to the classical equations of motion,
\begin{align}
  \label{eq:classical-eom}
  \nabla^2 \phicl_i(x) = \left[\phicl_j(x), \left[\phicl_j(x), \phicl_i(x)\right]\right].
\end{align}

For system I (c.f.\ table~\ref{Configurations}), the relevant solution to~\eqref{eq:classical-eom} with $\grp{SO}(3) \times \grp{SO}(3)$ symmetry is~\cite{Kristjansen:2012tn}
\begin{align}
  \label{eq:vevs-so3-so3}
  \begin{split}
    \varphicl_i(x) &= -\frac{1}{x_3}
    \begin{pmatrix}
      t_i^{(k_1)} \otimes \idm^{(k_2)} & 0 \\
      0 & 0^{(N - k_1 k_2)}
    \end{pmatrix}, \; i = 1, 2, 3, \\
    \varphicl_i(x) &= -\frac{1}{x_3}
    \begin{pmatrix}
      \idm^{(k_1)} \otimes \,\,t_{i - 3}^{(k_2)} & 0 \\
      0 & 0^{(N - k_1 k_2)}
    \end{pmatrix}, \; i = 4, 5, 6.
  \end{split}
\end{align}
Here the matrices $t_i^{(k)}$ constitute the $k$-dimensional irreducible representation of the Lie algebra $\alg{su}(2)$ and we denote by $0^{(N - k_1 k_2)}$ the zero matrix of dimension $(N - k_1 k_2) \times (N - k_1 k_2)$.
We will only need the explicit form of the diagonal matrix $t_{3}^{(k)}$; its eigenvalues are
\begin{align}
  d_{j,k} = \frac{1}{2} (k - 2j + 1), \hspace{0.5cm} j=1,\ldots, k.
\end{align}

For system II (c.f.\ table~\ref{Configurations}), the solution to~\eqref{eq:classical-eom} with $\grp{SO}(5)$ symmetry is given by~\cite{Constable:2001ag,Castelino:1997rv}
\begin{align}
  \label{eq:vevs-so5}
  \phicl_i(x) = \frac{1}{\sqrt{2} x_3}
  \begin{pmatrix}
    G_{i6} & 0 \\
    0 & 0^{(N - d_G)}
  \end{pmatrix}, \; i = 1, \ldots 5;
        \quad
        \phicl_6(x) = 0.
\end{align}
The matrices $G_{i6}$ together with $G_{ij} = -i \left[G_{i6}, G_{j6}\right]$ form the $d_G = \frac{1}{6} (n + 1) (n + 2) (n + 3)$ dimensional irreducible representation of the Lie algebra of $\grp{SO}(6)$.
For the purpose of this paper, we only need an explicit representation of $G_{56}$.
This matrix can be taken to be diagonal~\cite{deLeeuw:2016ofj}; its eigenvalues $\eta_{j,n}$ and the corresponding multiplicity $\mu_{j,n}$ are
\begin{align}
  \eta_{j,n} = -\frac{n}{2} + j - 1,
  \quad
  \mu_{j,n} = j (n - j + 2), \quad j=1,\ldots, n+1.
\end{align}

Note that for both systems, the classical solutions~\eqref{eq:vevs-so3-so3} and~\eqref{eq:vevs-so5} pertain to $x_3 > 0$.
The vevs for all other fields in $\mathcal{N} = 4$ SYM are zero in this region.
For $x_3 < 0$, the vevs for all fields vanish.

We shall calculate the expectation value of the Wilson line perturbatively in $\lambda$ at tree level and at one-loop, and in both cases consider only the leading order
in respectively $n$ and $k_1, k_2$ as $n,k_1,k_2 \rightarrow \infty$. This is motivated by a string theory analysis~\cite{Nagasaki:2012re,Kristjansen:2012tn}, which 
introduced the following double scaling limits (d.s.l.)
\begin{align}
\text{I}&:\quad  \lambda\rightarrow\infty, \quad     k_1,\;k_2\rightarrow\infty,  \quad \frac{\lambda}{\pi^2 \left(k_1^2+k_2^2\right)}\;\; \text{finite},\label{dslI}\\
\text{II}&:\quad \lambda\rightarrow\infty, \quad     n\rightarrow\infty,\quad \frac{\lambda}{\pi^2\;n^2}\;\; \text{finite}\,,\label{dslII}
\end{align}
where in case I also the ratio $k_2/k_1$ has to be taken finite.  Imposing the d.s.l. on the string theory side allows one to expand string theory observables,
such the expectation value of the Wilson line, as a power series in the double scaling parameter and formally compare the result to a perturbative gauge theory computation.

\subsection{Wilson line setup}
\label{sec:wilson-line-setup}

\begin{figure}
  \centering
  \def\svgwidth{\columnwidth}
  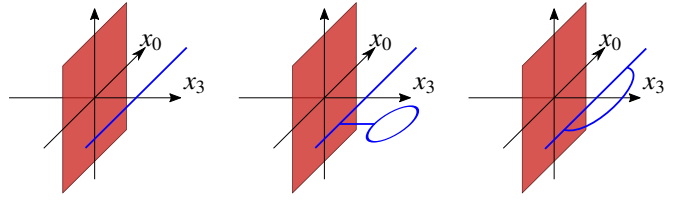
  \caption{Diagrams at tree level and one-loop order. (Figure adapted from \cite{deLeeuw:2016vgp}.)}
  \label{fig:diagrams}
\end{figure}

As in~\cite{Nagasaki:2011ue,deLeeuw:2016vgp}, we consider a straight
Wilson line parallel to the defect parametrized by $\gamma(t) = (t, 0, 0, z)$, i.e.\ a straight line at a fixed distance $z$ from the defect. For this case, the Wilson line is given by
\begin{align}
    \label{eq:parallel-propagator}
  \tr U(\alpha,\beta) = \tr\left[\pexp \int_{\alpha}^{\beta} dt\, \calA(t)\right],
\end{align}
with
\begin{align}
  \label{eq:curly-A-I}
  \calA^{(\mathrm{I})}(t) &= i A_0(t) - \varphi_3(t) \sin (\chi) - \varphi_6(t) \cos (\chi), \\
  \label{eq:curly-A-II}
  \calA^{(\mathrm{II})}(t) &= i A_0(t) - \phi_5(t) \sin (\chi) - \phi_6(t) \cos (\chi),
\end{align}
for the two set-ups respectively. We will be interested in the gauge invariant infinite line given by 
\begin{align}
\label{eq:wilsonline-limit}
W=\lim_{T\rightarrow\infty} \tr{} U\left(-\frac{T}{2},\frac{T}{2}\right),
\end{align}
which is related to the physical particle-interface potential. In order to compute the expectation value of the Wilson line, we expand the fields around the classical solution as
\begin{align}
\label{eq:}
\calA(t) =\calA^{\mathrm{cl}}(t)+\tilde{\calA}(t). 
\end{align}
To one-loop order, the path-ordered exponential becomes
\begin{align}
  \label{eq:parallel-propagator-expansion}
  \begin{split}
    &U(\alpha, \beta) = \Ucl(\alpha, \beta) + \int_{\alpha}^{\beta} dt \, \Ucl(\alpha, t) \tilde{\calA}(t) \Ucl(t, \beta) \\
    &\, + \int_{\alpha}^{\beta} dt \int_{t}^{\beta} dt' \, \Ucl(\alpha, t) \tilde{\calA}(t) \Ucl(t, t') \tilde{\calA}(t') \Ucl(t', \beta) + \mathcal{O}\left(\tilde{\calA}^3\right),
  \end{split}
\end{align}
where $\Ucl(\alpha, \beta)$ is the path-ordered exponential for the classical solution. The corresponding diagrams are illustrated in Figure~\ref{fig:diagrams} and the following subsections will be devoted to dealing with each of the terms.

\subsection{Tree-level}
\label{sec:tree-level}
The tree level contribution is given by the first term of~\eqref{eq:parallel-propagator-expansion} and is now evaluated in the large $T$ limit,
\begin{align}
\label{eq:tree1}
  \langle W\rangle_{\mathrm{tree}}&=\lim_{T\rightarrow\infty}\tr{} \pexp \int_{-T/2}^{T/2} dt\, \calA^{\mathrm{cl}}(t)\\
  &=\lim_{T\rightarrow\infty} \left[ \exp \left(T \calA^{\mathrm{cl}}\right)\right]_{i,i},
\end{align}
since the classical solutions are time-independent.
In the large $T$ limit, only the largest eigenvalue of $\calA^{\mathrm{cl}}$ contributes, which gives
\begin{align}
\label{eq:tree2}
\langle W\rangle_{\mathrm{tree}}=\mu  \exp\left(T \frac{\eta}{z} \right),
\end{align}
where $\eta/z$ is the largest eigenvalue of $\calA^{\mathrm{cl}}$ and $\mu$ its multiplicity. 
For the first setup, we have $2\eta^{(\mathrm{I})}=(k_{1}-1)\sin(\chi)+(k_{2}-1)\cos(\chi)$ and  $\mu^{(\mathrm{I})}=1$.  
For the second setup, we have $\eta^{(\mathrm{II})}=\frac{n}{\sqrt{8}}\sin(\chi)$ and $\mu^{(\mathrm{II})}=(n+1)$.
We may thus write the tree level results as
\begin{align}
  \label{eq:w-tree}
  \langle W\rangle_{\mathrm{tree}}^{(\mathrm{I})}&=\mu^{(\mathrm{I})}  \exp\left(T \frac{(k_{1}-1)\sin(\chi)+(k_{2}-1)\cos(\chi)}{2 z} \right),\\
  \langle W \rangle_{\mathrm{tree}}^{(\mathrm{II})} &= \mu^{(\mathrm{II})} \exp\left(T\frac{n \sin (\chi)}{\sqrt{8} z} \right).
\end{align}
They lead to the following particle-interface potentials
\begin{align}
  \label{eq:potential-tree}
    V_{\mathrm{tree}}^{(\mathrm{I})} &= -\lim_{T \rightarrow \infty} \frac{1}{T}\log{} \langle W \rangle_{\mathrm{tree}}^{(\mathrm{I})}
  = -\frac{k_{1}\sin(\chi)+k_{2}\cos(\chi)}{2 z} ,\\
  V_{\mathrm{tree}}^{(\mathrm{II})} &= -\lim_{T \rightarrow \infty} \frac{1}{T}\log{} \langle W \rangle_{\mathrm{tree}}^{(\mathrm{II})}
  = -\frac{n \sin (\chi)}{\sqrt{8} z} ,
\end{align}
having taken the  limit $k_1,k_2\rightarrow \infty$ in \eqref{eq:potential-tree} as implied by the double scaling limit.

\subsection{Lollipop}
\label{sec:lollipop}

The focus of this subsection is the second term of~\eqref{eq:parallel-propagator-expansion}, which involves the one-loop expectation value of $\tilde{\cal A}$ and which
we call the lollipop contribution. 
\begin{align}
\label{eq:lol1}
  \langle W \rangle_{\mathrm{lol}} &=\lim_{T\rightarrow\infty}\left\langle\tr\int_{-T/2}^{T/2} dt \, \Ucl\left(-\tfrac{T}{2}, t\right) \tilde{\calA}(t) \Ucl\left(t, \tfrac{T}{2}\right)\right\rangle\\
   &=\lim_{T\rightarrow\infty} T \left[e^{T \calA^{\mathrm{cl}}}\right]_{ij}\left\langle \left[\tilde{\calA}\right]_{ji} \right\rangle_{\mathrm{1-loop}},
\end{align}
where we have used the fact that the expectation values are time independent. The one-loop corrections to the vevs for the two set-ups are given in \cite{Grau:2018keb,Gimenez-Grau:2019fld}.  Notice that as opposed to what was the case for the supersymmetric D3-D5 probe brane set-up~\cite{Buhl-Mortensen:2016pxs,Buhl-Mortensen:2016jqo}, these corrections are non-vanishing. 
In the large $T$ limit, only the components multiplying the fastest growing exponential will contribute, which in both conventions is also the first component
\begin{align}
\label{eq:lol2}
 \langle W \rangle_{\mathrm{lol}} &=   T \mu e^{T \eta/x_{3}}\left\langle \left[\tilde{\calA}\right]_{11} \right\rangle_{\mathrm{1-loop}}.
\end{align}
Given the one-loop correction to the vevs, we find
\begin{align}
  \label{eq:w-lollipop}
  \langle W \rangle_{\mathrm{lol}}^{(\mathrm{I})}&=-\mu^{(\mathrm{I})} \frac{\lambda Te^{T \eta^{\left(\mathrm{I}\right)}/ z}}{4 \pi^2 z \left(k_{1}^{2}+k_{2}^{2}\right)^{3}} \left( k_{1}k_{2}^{4}\sin(\chi)+k_{2}k_{1}^{4}\cos(\chi)  \right),\\
    \langle W \rangle_{\mathrm{lol}}^{(\mathrm{II})}   &= -\mu^{(\mathrm{II})} \frac{\lambda Te^{T \eta^{\left(\mathrm{II}\right)}/ z}}{4\sqrt{8} \pi^2 z n}   \sin(\chi) ,
\end{align}
having again taken the double scaling limit in \eqref{eq:w-lollipop}.

\subsection{Tadpole}
\label{sec:tadpole}
As in~\cite{deLeeuw:2016vgp}, the third term of~\eqref{eq:parallel-propagator-expansion} is the least straight forward term to compute. However, the same techniques can be employed with just minor complications. The tadpole term is
\begin{align}
  U_{\mathrm{tad}}(\alpha,\beta) &=    \int_{\alpha}^{\beta}dt\int_{t}^{\beta}dt' U^{\mathrm{cl}}(\alpha,t) \tilde{\mathcal{A}}(t)U^{\mathrm{cl}}(t,t' )  \tilde{\mathcal{A}}(t' )U^{\mathrm{cl}}(t' ,\beta).
\end{align}
The fields are all $N \times N$ matrices;
decomposing them into the block structure given by the classical solutions~\eqref{eq:vevs-so3-so3} and~\eqref{eq:vevs-so5} and writing out the matrix indices explicitly, we find
\begin{align}
\label{eq:tad2}
  &\langle\tr{} U_{\mathrm{tad}}(\alpha,\beta)\rangle =
    \int_{\alpha}^{\beta}dt\int_{t}^{\beta}dt'  \langle[\tilde{\mathcal{A}}(t)]_{\mu\rho} [ \tilde{\mathcal{A}}(t' )]_{\rho\mu}\rangle \\
  &\, \nonumber
    +\int_{\alpha}^{\beta}dt\int_{t}^{\beta}dt'  \left[e^{(t' -t) \mathcal{A}^{\mathrm{cl}}}\right]_{cd} \langle[ \tilde{\mathcal{A}}(t' )]_{d\mu}[\tilde{\mathcal{A}}(t)]_{\mu c}\rangle \\
  &\, \nonumber
    +\int_{\alpha}^{\beta}dt\int_{t}^{\beta}dt' \left[e^{(\beta-\alpha+t-t' ) \mathcal{A}^{\mathrm{cl}}}\right]_{eb} \langle[\tilde{\mathcal{A}}(t)]_{b\rho} [ \tilde{\mathcal{A}}(t' )]_{\rho e}\rangle \\
  &\, \nonumber
    +\int_{\alpha}^{\beta}dt\int_{t}^{\beta}dt' \left[e^{(\beta-\alpha+t-t' ) \mathcal{A}^{\mathrm{cl}}}\right]_{eb} \left[e^{(t' -t) \mathcal{A}^{\mathrm{cl}}}\right]_{cd} \langle[\tilde{\mathcal{A}}(t)]_{bc} [ \tilde{\mathcal{A}}(t' )]_{de}\rangle.
\end{align}
For the first setup the latin indices run from $1$ to $k_{1}k_{2}$ and the greek indices run from $k_{1}k_{2}+1$ to $N$, while for the second setup the latin indices run from $1$ to $d_{G}$ and the greek indices run from $d_{G}+1$ to $N$. In the large $N$ limit only the second and third term contribute, given the propagators found in~\cite{Grau:2018keb,Gimenez-Grau:2019fld}. We thus have
\begin{align}
\label{eq:tad3}
\langle W \rangle_{\mathrm{tad}} =&\lim_{T\rightarrow\infty}\int_{-T/2}^{T/2}d\alpha\int_{\alpha}^{T/2}d\beta \nonumber\\&\bigg[ e^{ -(\alpha-\beta)\mathcal{A}^{\mathrm{cl}} } 
  +e^{ (\alpha-\beta+T)\mathcal{A}^{\mathrm{cl}} } \bigg]_{cd} \langle [\tilde{\mathcal{A}}]_{d\mu}(\alpha)[\tilde{\mathcal{A}}]_{\mu c}(\beta)\rangle.
\end{align}
For both setups the propagator has the form
\begin{align}
\label{eq:tad4}
\langle [\tilde{\mathcal{A}}]_{d\mu}(\alpha)[\tilde{\mathcal{A}}]_{\mu c}(\beta)\rangle=\sum_{n}D^{n}_{dc}\sum_{i}\lambda_{i,n}K^{m^{2}_{i,n}} (\alpha, \beta),
\end{align}
where $D$ is a diagonal prefactor and $K^{m^{2}_{i,n}}$ is the spacetime part of the propagator given in~\eqref{eq:tad6} below. This means we have to perform integrals of the form
\begin{align}
\label{eq:tad5}
  &\langle W \rangle_{\mathrm{tad}} =\lim_{T\rightarrow\infty}\int_{-T/2}^{T/2}d\alpha\int_{\alpha}^{T/2}d\beta \nonumber\\
  &\quad \bigg[ e^{ -(\alpha-\beta)\mathcal{A}^{\mathrm{cl}} } 
  +e^{ (\alpha-\beta+T)\mathcal{A}^{\mathrm{cl}} } \bigg]_{cd} \sum_{n}D^{n}_{dc}\sum_{i}\lambda_{i,n}K^{m^{2}_{i,n}}(\alpha, \beta).
\end{align}
Following~\cite{deLeeuw:2016vgp}, we proceed by using the following representation of the propagator
\begin{align}
\label{eq:tad6}
  K^{m^{2}_{i}}(\alpha, \beta) &= \frac{g^2_{YM} z}{4\pi^2} \int_{0}^{\infty}dr \, r \frac{\sin(\delta r)}{\delta} I_{\nu_{i}}(r z)K_{\nu_{i}}(r z),\\
   \nu_{i}&=\sqrt{m^{2}_{i}+\frac{1}{4}},
\end{align}
having defined $\delta=\beta-\alpha$. We may now plug this back into \eqref{eq:tad5}, change variables $\alpha=\delta-T/2$, rescale $r\rightarrow r/z$ and do the $\beta$ integration,
\begin{align}
\label{eq:tad7}
\nonumber\langle W \rangle_{\mathrm{tad}} =&\frac{g^2_{YM}}{4\pi^2z}\lim_{T\rightarrow\infty}\int_{0}^{T}d\delta \, (T-\delta) \int_{0}^{\infty}dr \, r \frac{\sin(\delta r/z)}{\delta}\\&\bigg[ e^{ \delta\mathcal{A}^{\mathrm{cl}} } 
  +e^{ (T-\delta)\mathcal{A}^{\mathrm{cl}} } \bigg]_{cd} \sum_{n}D^{n}_{dc}\sum_{i}\lambda_{i,n} I_{\nu_{i,n}}(r )K_{\nu_{i,n}}(r).
\end{align}
Integration by parts is performed on the $r$ integration in order to cancel the $\frac{1}{\delta}$ such that the integration over $\delta$ can be carried out,
\begin{align}
\label{eq:tad8}
  &\langle W \rangle_{\mathrm{tad}} =
    \frac{g^2_{YM}}{4\pi^2z^{2}}\lim_{T\rightarrow\infty}\int_{0}^{T}d\delta\, (T-\delta)\bigg[ e^{ \delta\mathcal{A}^{\mathrm{cl}} } 
    +e^{ (T-\delta)\mathcal{A}^{\mathrm{cl}} } \bigg]_{cd} \\
  &\quad \nonumber
    \sum_{n}D^{n}_{dc} \int_{0}^{\infty}dr  \cos(\delta r/z)\int_{r}^{\infty} dr'r' \sum_{i}\lambda_{i,n} I_{\nu_{i,n}}(r'  )K_{\nu_{i,n}}(r' ).
\end{align}
Using this antiderivative makes the boundary term vanish at infinity, whilst the $\sin(\delta r/z)$ part makes the boundary term vanish at $r=0$. We can now perform the $\delta$ integration. In the large $T$ limit we have
\begin{align}
\label{eq:tad9}
  &\int_{0}^{T}d\delta \, (T-\delta) \bigg[ e^{ \delta\mathcal{A}^{\mathrm{cl}} } \nonumber
    +e^{ (T-\delta)\mathcal{A}^{\mathrm{cl}} } \bigg]_{cd} \sum_{n}D^{n}_{dc}   \cos(\delta r/z) \\
  &\qquad = \mu e^{\eta T/z} T z\, \frac{\eta}{\eta^{2}+r^{2}}\, \sum_{n}D^{n}_{1,1},
\end{align}
since for our two setups the largest eigenvalue of $D$ coincides with the largest eigenvalue of $\mathcal{A}^{\mathrm{cl}}$. We use this result in \eqref{eq:tad8},
\begin{align}
\label{eq:tad10}
\nonumber\langle W \rangle_{\mathrm{tad}} =\mu&\frac{g^2_{YM} Te^{\eta T/z}  }{4\pi^2z} \int_{0}^{\infty}dr \, \frac{\eta}{\eta^{2}+r^{2}}\\&\sum_{n}D^{n}_{1,1}\int_{r}^{\infty}dr'\, r' \sum_{i}\lambda_{i,n} I_{\nu_{i,n}}(r'  )K_{\nu_{i,n}}(r' ).
\end{align}
It is here and in the following implicit that $T$ is large. We will now perform the $r' $ integration in the double scaling limit and for convenience we define the functions $A$ and $F$
\begin{align}
\label{eq:tad11}
  A(r)&=\int_{r}^{\infty}dr' \, r' \sum_{i}\lambda_{i,n} I_{\nu_{i,n}}(r'  )K_{\nu_{i,n}}(r' )\\&=-\sum_{i}\lambda_{i,n}F_{\nu_{i,n}}(r)+\lim_{r' \rightarrow\infty}\sum_{i}\lambda_{i,n}F_{\nu_{i,n}}(r' ),\\\label{eq:tad12}
  F_{\nu_{i,n}}(r)&=\int_{0}^{r} dr' \, r' I_{\nu_{i,n}}(r'  )K_{\nu_{i,n}}(r' ).
\end{align}
By doing the integral from~\eqref{eq:tad12}, we find $F_{\nu_{i,n}}(r)$ to be
\begin{align}
\label{eq:tad13}
  F_{\nu_{i,n}}(r)=-\frac{\nu_{i,n}}{2}+\frac{1}{2} \left(r^{2}+\nu^{2}_{i,n}\right) I_{\nu_{i,n}}(r)K_{\nu_{i,n}}(r)-\frac{1}{2} r^{2}I^{\prime}_{\nu_{i,n}}(r)K^{\prime}_{\nu_{i,n}}(r).
\end{align}
In the double scaling limit, we can use the behaviour of the Bessel functions at large order and finite argument~\cite{NIST} and find
\begin{align}
\label{eq:tad14}
F_{\nu_{i,n}}(r)&=-\frac{\nu_{i,n}}{2}+\frac{1}{2} \left(\nu_{i,n}^{2}+r^{2}\right)^{1/2} + \mathcal{O}\left(\nu_{i,n}^{-1}\right).
\end{align}
We note that $A(r)$ is divergent unless $\sum_{i}\lambda_{i,n}=0$, but by properly bunching our terms we can show that this condition is satisfied.
Then, we find
\begin{align}
\label{eq:tad15}
A(r)=-\frac{1}{2}\sum_{i} \lambda_{i,n}  \left(\nu_{i,n}^{2}+r^{2}\right)^{1/2} + \mathcal{O}\left(\nu_{i,n}^{-1}\right).
\end{align}
This result is now plugged into \eqref{eq:tad10} and the final integral is performed
\begin{align}
\label{eq:tad16}
  \langle W \rangle_{\mathrm{tad}}
  &=-\mu\frac{g^2_{YM} Te^{\eta T/z}  }{8\pi^2z}\sum_{n}D^{n}_{1,1} \\
  &\hspace{1cm} \nonumber \int_{0}^{\infty}dr \, \frac{\eta}{\eta^{2}+r^{2}}\sum_{i} \lambda_{i,n}  \left(\nu_{i,n}^{2}+r^{2}\right)^{1/2} \\
  &=-\mu\frac{g^2_{YM} Te^{\eta T/z}  }{16\pi^2z}\sum_{n}D^{n}_{1,1} \\
  &\hspace{0.2cm} \nonumber \sum_{i} \lambda_{i,n}  \left[2\sqrt{\nu_{i,n}^{2}-\eta^{2}}\  \mathrm{arccot} \left(\tfrac{\eta}{\sqrt{\nu_{i,n}^{2}-\eta^{2} }}\right)-\eta \log\left(\nu_{i,n}^{2}\right) \right],
\end{align}
where we again used $\sum_{i}\lambda_{i,n}=0$ in the second line.
We finally plug in the coefficients for the first setup, let $k_{2}=k_{1}\tan(\psi_{0})$ and take the large $k_{1}$ limit
\begin{align}
\label{eq:tad17}
\nonumber  \langle W\rangle_{\mathrm{tad}}^{(\mathrm{I})}=-\mu^{(\mathrm{I})}&\frac{\lambda T e^{\nicefrac{T\eta^{(\mathrm{I})}}{z} }\cos(\psi_{0})}{4\pi^{2}z\,k_{1}}  \frac{\sin^{2}(\psi_{0}+\chi)}{4\cos^{3}(\psi_{0}+\chi)}\\&\hspace{1cm}\left(2\psi_{0}+2\chi-\pi+\sin(2\psi_{0}+2\chi)\right),
\end{align}
notice that $\cos(\psi_{0})/k_{1}=\left(k_{1}^{2}+k_{2}^{2}\right)^{-1/2}$ gives the combination appearing in the double scaling parameter. For the second setup in the large $n$ limit we find
\begin{align}
\label{eq:tad18}
  \langle W\rangle_{\mathrm{tad}}^{(\mathrm{II})}=-\mu^{(\mathrm{II})}&\frac{ \lambda Te^{{\nicefrac{T\eta^{(\mathrm{II})}}{z} } }}{4\sqrt{8}\pi^2z n} \frac{\sin^{2}(\chi)}{\cos^{3}(\chi)}\left(2\chi-\pi+\sin(2\chi)  \right) .
\end{align}

\subsection{Full one-loop result}
\label{sec:one-loop}
The full one-loop result is now obtained by adding the lollipop and the tadpole contribution
\begin{align}
\label{eq:oneloop1}
\nonumber  \langle W\rangle_{\mathrm{1-loop}}^{(\mathrm{I})}=&-\mu^{(\mathrm{I})}\frac{\lambda T e^{\nicefrac{T\eta^{(\mathrm{I})}}{z} }\cos(\psi_{0})}{4\pi^{2}zk_{1}}  \bigg[ \cos(\chi)  \sin(\psi_{0})\cos^{4}(\psi_0)\\&\nonumber+\sin(\chi) \cos(\psi_{0})\sin^{4}(\psi_{0}) \\&+\frac{\sin^{2}(\psi_{0}+\chi)}{4\cos^{3}(\psi_{0}+\chi)}\left(2\psi_{0}+2\chi-\pi+\sin(2\psi_{0}+2\chi)\right) \bigg],
\end{align}
having also expressed the lollipop contribution in terms of $ \psi_{0}=\arctan(k_{2}/k_{1})$. For the second setup we have
\begin{align}
  \label{eq:w-one-loop}
  \nonumber\langle W \rangle_{\mathrm{1-loop}}^{(\mathrm{II})} &= -\mu^{(\mathrm{II})} \frac{\lambda T}{4 \pi^2 n} \frac{e^{\nicefrac{T\eta^{(\mathrm{II})}}{z}}}{\sqrt{8} z}\\& \qquad \bigg[\sin (\chi) - \frac{\sin^2 (\chi)}{\cos^3 (\chi)}\left(\pi - 2 \chi - \sin(2 \chi)\right)\bigg].
\end{align}
The corresponding correction to the particle-interface potential is given by
\begin{align}
\label{eq:1}
V_{\mathrm{1-loop}}=-\lim_{T\rightarrow\infty} \frac{1}{T} \frac{\langle W \rangle_{\mathrm{1-loop}}}{\langle W \rangle_{\mathrm{tree}}},
\end{align}
which concludes the gauge theory computation with the following results:
\begin{align}
  \label{eq:potential-one-loop}
  V_{\mathrm{1-loop}}^{(\mathrm{I})}
  &= V_{\mathrm{tree}}^{(\mathrm{I})} \, \left(\frac{\lambda}{\pi^2 \left(k_1^2 + k_2^2\right)}\right) \, \frac{1}{2 \sin(\psi_0 + \chi)} \\
  &\quad \Bigg[\frac{\sin^2(\psi_0 + \chi)}{4 \cos^3(\psi_0 + \chi)} \left(\pi-2 \psi_0 - 2 \chi -  \sin(2 \psi_0 + 2 \chi)\right) \nonumber\\
  &\quad - \cos(\chi) \sin(\psi_0) \cos^4(\psi_0) - \sin(\chi) \cos(\psi_0) \sin^4(\psi_0) \nonumber \Bigg], \nonumber \\
  V_{\mathrm{1-loop}}^{(\mathrm{II})}
  &= V_{\mathrm{tree}}^{(\mathrm{II})} \, \left(\frac{\lambda}{\pi^2 n^2}\right) \, \left[\frac{\sin(\chi)}{4 \cos^3(\chi)} \left(\pi - 2 \chi - \sin(2 \chi)\right) - \frac{1}{4}\right].
\end{align}

\section{The String Theory Computation}
\label{sec:string-theory-computation}

 As summarized in table~\ref{Configurations}, we will be considering two different D3-D7 probe brane systems.  In the set-up I, the probe D7-brane has geometry $AdS_4\times S^2\times S^2$, and a background gauge field has has $k_1$ and $k_2$ units of magnetic flux through the two $S^2$ spheres, respectively.
 In the second configuration, II, the D3-branes are intersected by a (small) number of  D7-branes with  $AdS_4\times S^4$ geometry and a background gauge field supports a non-vanishing instanton number on  $S^4$.  In both cases, the system is stabilized for sufficiently large values of the flux or instanton number.~\footnote{We notice that the perturbative
 regime for the double scaling parameter, considered in the gauge theory computations, lies within the region of stability of the probe brane systems~\cite{Grau:2018keb,Gimenez-Grau:2019fld}.}
 
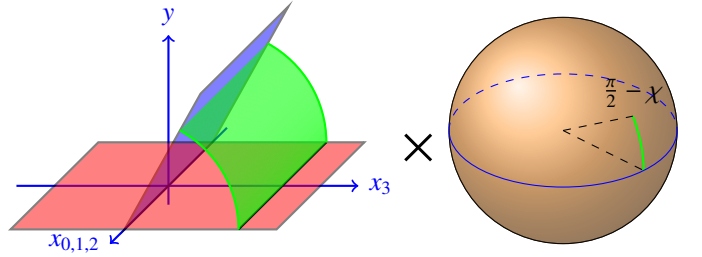
\begin{figure}[tb]
$
 \begin{aligned}
 \begin{tikzpicture}
 	[	D3/.style={opacity=.5,thick,fill=red},
 		D5/.style={opacity=.5,thick,fill=blue},
 		axis/.style={->,blue,thick},
 		string/.style={green,thick},
 		axisline/.style={blue,thick},
 		Wline/.style={black,thick},
 		cube/.style={opacity=.7, thick,fill=red}]

 	\draw[axisline] (-2,0,0) -- (0,0,0) node[anchor=west]{};	
			
 	\draw[axis] (0,0,0) -- (2.5,0,0) node[anchor=west]{$x_3$};
 	\draw[axis] (0,-0.25,0) -- (0,2,0) node[anchor=south]{$y$};
 	\draw[axis] (0,0,-2) -- (0,0,2) node[anchor=east]{$x_{0,1,2}$};

 	\draw[D5] (0,0,-1.5) -- (1,+1.8,-1.5) -- (1,1.8,1.5) -- (0,0,1.5) -- cycle;
 	
 	\draw[D3] (-1.5,0,-1.5) -- (2,0,-1.5) -- (2,0,1.5) -- (-1.5,0,1.5) -- cycle;
	
	\draw[Wline] (1.5,0,-1.5) -- (1.5,0,1.5) node[anchor=west]{};	
 	
\foreach \x in {-1.5,-1.49,...,1.5} 	{

     \tdplotdrawarc[string,opacity=0.15]{(0,0,\x)}{1.5}{61}{0}{}{};
}
     \tdplotdrawarc[string]{(0,0,-1.5)}{1.5}{61}{0}{}{};
     \tdplotdrawarc[string]{(0,0,1.5)}{1.5}{61}{0}{}{};
 
 \end{tikzpicture}
 \end{aligned} 
 \scalebox{2}{\raisebox{-0.3\baselineskip}{$\times$}}
 \,\,
\begin{aligned}
\tdplotsetmaincoords{60}{135}
\begin{tikzpicture}[scale=1.5,tdplot_main_coords]
\tikzstyle{string}=[thick,color=green,tdplot_rotated_coords]
     \tdplotsetrotatedcoords{0}{-90}{0}; 
    \draw[tdplot_main_coords] (0,0) circle (1cm);
    \draw[ball color=orange!80!white,opacity=0.69,tdplot_main_coords] (0,0) circle (1cm);
     \tdplotsetrotatedcoords{0}{-90}{0};
     \tdplotdrawarc[string]{(0,0,0)}{1}{60}{90}{}{};
      \draw[tdplot_main_coords,blue] (-1cm,0cm) arc (180:360:1cm and 0.5cm);
     \draw[dashed] (0,0,0) -- (0,1,0);
     \draw[dashed] (0,0,0) -- (0,0.86602540378,0.5) node[anchor=south] {$\frac{\pi}{2}-\chi$};
      \draw[dashed,tdplot_main_coords,blue] (-1cm,0cm) arc (180:0:1cm and 0.5cm);
\end{tikzpicture}
\end{aligned}
$
\caption{\label{fig: string theory} The minimal surface corresponding to the Wilson line for the set-up II. In the $AdS_5$ factor, the minimal surface (green) stretches from the Wilson line (black) on the boundary (red) to the D7 brane (blue). In the $S^5$ factor, it  stretches from the $S^4$ wrapped by the D7-brane (blue) along the perpendicular direction for
an angular extent of  $\frac{\pi}{2}-\chi$. For the set-up II, the minimal surface in the $AdS$ part of the geometry looks similar whereas in the spherical part it is somewhat different, cf.\ eqns.\ (\ref{metric1}), (\ref{metric2}). (Figure adapted from~\cite{deLeeuw:2016vgp}.)}
\end{figure}

It is convenient to write the $AdS_5 \times S^5$ metric in two different ways, depending on the D7 geometry that we are considering
\begin{align}
\label{metric1}
\text{d}s_{\text{I}}^2&=\frac{1}{y^2}\left( \text{d}y^2+\text{d}x_{\mu}\text{d}x_{\nu}\eta^{\mu \nu}\right) +\text{d}\psi^2+\cos^2 \psi \,\text{d} \Omega^2_{S^2}+\sin^2 \psi \,\text{d} \tilde{\Omega}^2_{S^2}, \\
\label{metric2}
\text{d}s_{\text{II}}^2&=\frac{1}{y^2}\left( \text{d}y^2+\text{d}x_{\mu}\text{d}x_{\nu}\eta^{\mu \nu}\right) +\text{d}\psi^2+\cos^2 \psi \,\text{d} \Omega^2_{S^4}\,,
\end{align}
where $\text{d} \Omega^2_{S^2}$ and $\text{d} \tilde{\Omega}^2_{S^2}$ are the metrics of the two $S^2$ spheres and $d\Omega^2_{S^4}$ denotes the metric of the $S^4$ inside the $S5$. In both cases $x_{\mu}=(x_0, x_1, x_2,x_3)$ and the boundary of $AdS_5$ is located at $y=0$. In the set-up I, the D7-brane has world volume coordinates $(x_0,x_1,x_2,y,\Omega_{S^2},\tilde{\Omega}_{S^2})$, while in the configuration II the D7-branes wrap the four-sphere and extend in the $( x_0,\,x_1,\,x_2,\,y) $ directions. The embedding of the D7 in the target space is given by \cite{Constable:2001ag,Myers:2008me,Kristjansen:2012tn}
\begin{equation}
\label{caseI}
\text{I:}\quad  y=\frac{x_3}{\Lambda_\text{I}}, \quad \Lambda_\text{I} = \frac{f_1f_2}{\sqrt{(f_1^2+4\cos^4\psi)(f_2^2+4\sin^4\psi)-f_1^2f_2^2}},
\end{equation}
where $f_{1,2}=\frac{2 \pi k_{1,2}}{\sqrt{\lambda}}$ and the angle $\psi$ has to satisfy
\begin{equation}
\label{angle}
(f_1^2+4\cos^4\psi)\tan^2\psi=(f_2^2+4\sin^4\psi)\,,
\end{equation}
\begin{align}
\label{caseII}
\text{II:} \quad y&=\frac{x_3}{\Lambda_\text{II}}, \quad\quad \psi=0, \\ &\text{when}\quad n \rightarrow \infty: \quad \Lambda_{\text{II}}\sim \frac{\pi n}{\sqrt{2}\sqrt{\lambda}}\!-\!\frac{\sqrt{\lambda}}{4\sqrt{2}\pi n}+ \mathcal{O}\left( \frac{\lambda^{3/2}}{\pi^3 n^3}\right). \nonumber
\end{align}
In both cases, the D7-brane intersects $AdS_5$ along an $AdS_4$ hyperplane, tilted with respect to the boundary $y = 0$ at
an angle that depends on $\Lambda_{\text{I}}$ or $\Lambda_{\text{II}}$.
In the supergravity limit $\lambda \rightarrow \infty$, following the idea of \cite{Rey:1998ik,Maldacena:1998im,Drukker:1999zq}, the Wilson line expectation value is described by the area of a minimal surface stretching from the boundary of $AdS_5$ to the D7-brane in the interior. Notice that the minimal surface attaches to the D7-brane along a straight line in its $AdS$ part as well as along an arc in its spherical part, cf.\ figure~\ref{fig: string theory}. 

We parametrise the worldsheet using coordinates $(\tau, \sigma)$ with $\tau \in [-\frac{T}{2},\frac{T}{2}] $ and $\sigma \in[0,\tilde{\sigma}]$. For the straight Wilson line (parallel to the defect) we make the following ansatz for the embedding of the string \cite{Nagasaki:2011ue,Preti:2017fhw}
\begin{equation}
\label{ansatz}
t=\tau,\quad\quad y=y(\sigma),\quad\quad x_3=x_3(\sigma)\quad\mathrm{and}\quad \psi=\psi(\sigma).
\end{equation}
A new feature in the defect set-up is that the extremal surface has to satisfy two different sets of boundary conditions. At the boundary of $AdS_5$, which is approached when $\sigma\to 0$, the usual Dirichlet boundary conditions must be imposed
\begin{equation}
y(0)=0,\quad\quad x_3(0)=z \quad \mathrm{and}\quad \psi(0)=\frac{\pi}{2}-\chi\,.
\end{equation}
 The second set of boundary conditions ensures that the extremal surface intersects  the boundary brane at  $\tilde{\sigma}$ orthogonally
\begin{align}
	\label{bc1}
\text{I:}\quad	y(\tilde\sigma)&=\frac{x_3(\tilde\sigma)}{\Lambda_{\text{I}}},\; y^\prime (\tilde\sigma)+{\Lambda_{\text{I}}} x^\prime_3(\tilde\sigma)=0, \; 
\psi(\tilde\sigma)=\psi_1, \\
	\label{bc2}
\text{II:}\quad	y(\tilde\sigma)&=\frac{x_3(\tilde\sigma)}{\Lambda_{\text{II}}},\; y^\prime (\tilde\sigma)+{\Lambda_{\text{II}}} x^\prime_3(\tilde\sigma)=0, \; 
	\psi(\tilde\sigma)=0\,,
\end{align}
where $\tilde \sigma$ is the maximum value of the worldsheet coordinate $\sigma$ and $\psi_1$ has to satisfy eq.~\eqref{angle} and $\psi_1 \in[0,\pi/2]$. The construction of the 
solution follows the idea of  \cite{Nagasaki:2011ue}. The Euclidean Polyakov action in conformal gauge reduces to 
\begin{equation}
\label{action}
S=\frac{1}{4\pi \alpha^{'}}\int d\tau d\sigma \, \frac{1}{y^2} \left(1+y^{\prime 2}+x_3^{\prime 2}+
y^2 \psi^{\prime 2}\right)\,.
\end{equation}
The Euler-Lagrange equations of motion for the action \eqref{action} must be combined with the Virasoro constraint 
\begin{equation}
y^{\prime 2}+x_3^{\prime  2}+
y^2 \psi^{\prime 2}=1.
\end{equation}
Since the coordinates $x_3$ and $\psi$ are cyclic variables, cf.\ \eqref{action}, their equations of motion immediately translate into two conservation laws
\begin{equation}
\label{conslaw}
x_3^\prime(\sigma)=-c y^2(\sigma)\quad\quad\mathrm{and}\quad\quad \psi^\prime(\sigma)=j,
\end{equation}
where $j$ and $c$ are two integration constants to be determined. The equation of motion for $y(\sigma)$ is given by
\begin{equation}
\label{diffy}
yy^{\prime\prime}-y^{\prime 2}+1+c^2 y^4=0\,.
\end{equation}
 Using the Virasoro constraint we get the following first order differential equation for $y^{\prime}(\sigma)$
\begin{equation}
\label{yp}
y^{\prime}=\sqrt{1-j^2y^2-c^2y^4}\,.
\end{equation}
The solutions to eqs. \eqref{conslaw} and \eqref{yp} are\footnote{Our notation for elliptic functions and integrals follows that of the Wolfram Language of Mathematica.}
\begin{align}
	y(\sigma)&=\sqrt{\frac{m+1}{j^2}}\text{sn}\left(\left.\sqrt{\frac{j^2}{m+1}}\sigma\right| m \right), \\
	x_3(\sigma)&=z\!-\!\!\sqrt{-\frac{m+1}{m\; j^2}}\left[\!\mathds{E}\left(\! \text{am}\left(\!\! \sqrt{\frac{j^2}{m+1}}\sigma,m\right)\!,m\right) \!-\!\! \sqrt{\frac{j^2}{m+1}}\sigma\right],\\
	\psi(\sigma)&=j\sigma+\frac{\pi}{2}-\chi\,,
\end{align}
where to determine the form of the solutions we have used the boundary conditions at $\sigma=0$. The parameter $m$ is the elliptic modulus and it ranges from $0$ to $-1$.The boundary conditions on $\tilde{\sigma}$ fix the remaning parameters $(\tilde{\sigma},\;j,\;m)$ in terms of the geometrical data $(z,\;\Lambda_{\text{I}} \text{ or }\Lambda_{\text{II}},\;\chi)$
\begin{align}
\label{sigmaI}
\text{I}: \quad\tilde \sigma&=\frac{1}{j}\left(\psi_1+\chi-\frac{\pi}{2}\right), \qquad \qquad  0\leq \psi_1+\chi \leq \frac{\pi}{2},\\
\label{sigmaII}
\text{II}: \quad \tilde \sigma&=\frac{1}{j}\left(\chi-\frac{\pi}{2}\right), \qquad \qquad \qquad \qquad  0\leq \chi \leq \frac{\pi}{2},\\
\label{j}
j^2&=\left( \frac{\Lambda}{z}\sqrt{m+1}\,\text{sn}\left(\left. \sqrt{\frac{j^2}{m+1}}\tilde{\sigma}\right| m\right) +\right.\nonumber\\&\left.+\frac{\sqrt{m+1}}{z\sqrt{-m}}\left[\mathds{E}\left( \text{am}\left( \sqrt{\frac{j^2}{m+1}}\tilde{\sigma},m\right) ,m\right)\!-\!\!\sqrt{\frac{j^2}{m+1}}\tilde{\sigma} \right]\;\right) ^2, \,\\
\label{flux}
 \Lambda_{\text{I},\;(\text{II})}&=\frac{\text{cn}\left(\left. \sqrt{\frac{j^2}{m+1}}\tilde{\sigma}_{\text{I},\;(\text{II})}\right| m\right)\text{dn}\left(\left. \sqrt{\frac{j^2}{m+1}}\tilde{\sigma}_{\text{I},\;(\text{II})}\right| m\right)}{\sqrt{-m}\;\text{sn}\left(\left. \sqrt{\frac{j^2}{m+1}}\tilde{\sigma}_{\text{I},\;(\text{II})}\right| m\right)^2}\,.
\end{align}
Choosing the convention in which $\tilde{\sigma}$ is positive, for the value of the angles considered in \eqref{sigmaI} and \eqref{sigmaII}, $j$ has to be negative.\\
The area of the minimal surface is obtained by evaluating the Polyakov action on the classical solution. As usual, one has to introduce a cut-off $\epsilon$ in the $y$ coordinate such that the regularized area is given by an integral in the region $y\geq \epsilon$ and then remove the divergent piece before comparing to the field-theory computation. The expression for the regularized action is
\begin{align}
\label{ren}
\mathcal{S}_{\text{I},\;(\text{II})}&=\frac{\sqrt{\lambda}T }{2\pi}\sqrt{\frac{j^2}{m+1}}\left[ 
\sqrt{\frac{j^2}{m+1}}\tilde{\sigma}_{\text{I},\;(\text{II})}-\nonumber\right.\\&-\left.\mathds{E}\left(\text{am}\left(\left.\left.\sqrt{\frac{j^2}{m+1}}\tilde{\sigma}_{\text{I},\;(\text{II})}
		\,\right |m\right)\,\right|m\right)-\nonumber\right.\\&-\left.\frac{\text{cn}\left( \left.\sqrt{\frac{j^2}{m+1}}\tilde{\sigma}_{\text{I},\;(\text{II})}\,\right| m\right) \text{dn}\left(\left. \sqrt{\frac{j^2}{m+1}}\tilde{\sigma}_{\text{I},\;(\text{II})}\right| m\right)}{\text{sn}\left(\left. \sqrt{\frac{j^2}{m+1}}\tilde{\sigma}_{\text{I},\;(\text{II})}\right| m\right)}\right].
\end{align}
We can rewrite $\mathcal{S}_{\text{I},\;(\text{II})}$ in a more compact form using eq.~ \eqref{j} to replace the incomplete elliptic integral of the second kind and noticing that
\begin{equation}
y^{\prime}(\tilde{\sigma})=\text{cn}\left(\left. \sqrt{\frac{j^2}{m+1}}\tilde{\sigma}_{\text{I},\;(\text{II})}\right| m\right) \text{dn}\left(\left. \sqrt{\frac{j^2}{m+1}}\tilde{\sigma}_{\text{I},\;(\text{II})}\right| m\right)\,,
\end{equation}
we get
\begin{equation}
\label{ren2}
\mathcal{S}_{\text{I},\;(\text{II})}=-\frac{\sqrt{\lambda}T }{2\pi}z\,c\,,
\end{equation} 
where $c=\frac{j^2\,\sqrt{-m}}{m+1}$. 
To compare the supergravity and the gauge theory results, we have to expand our results in the double scaling parameter given in eqns.~(\ref{dslI}) and~(\ref{dslII}).
One can get the expansion for $\Lambda_{\text{I}}$ in powers of $\frac{\lambda}{\pi^2(k_1^2+k_2^2)}$ looking at its definition in eq.~\eqref{caseI}. Notice that eq.~\eqref{angle} is satisfied in the large flux limit if
\begin{equation}
\psi_1=\psi_0+\frac{\cos \psi_0(\sin \psi_0-\sin 3 \psi_0)}{4 \pi^2}\frac{\lambda}{k_1^2+k_2^2}+\mathcal{O}\left( \frac{\lambda^2}{\pi^4\left( k_1^2+k_2^2\right) ^2}\right) \,,
\end{equation}
where $\tan \psi_0 =\frac{k_2}{k_1}$.
Thus, the expansion for $\Lambda_{\text{I}}$ is
\begin{equation}
\Lambda_{\text{I}}= \sqrt{\frac{k_1^2+k_2^2}{\lambda}}\pi-\frac{\sin^2 2\psi_0}{8\pi}\sqrt{\frac{\lambda}{k_1^2+k_2^2}}+\mathcal{O}\left( \frac{\lambda}{\pi^2(k_1^2+k_2^2)}\right)\,. 
\end{equation} 
The double-scaling expansion for $\Lambda_{\text{II}}$ can be read off from eq.~\eqref{caseII}. Notice that in this limit also $\Lambda_{\text{I,II}}$ have to be large. Namely, we require that the denominator in eq.~\eqref{flux} vanishes. This occurs when $m$ goes to zero. Moreover, we can assume the following expansion for $m$
\begin{equation}
m=\sum_{l=1}^{\infty}\frac{a_{2l}}{\Lambda^{2l}_{\text{I,II}}},
\end{equation}
in such a way that eq.~\eqref{flux} is satisfied. The coefficient in the above expansion can be determined by solving iteratively equation \eqref{flux}. In the end, we get the following expansions for the particle-defect potential in the two different cases I and II
\begin{align}
  \label{actionI}
  V^{(\mathrm{I})}
  &= -\frac{k_1 \sin(\chi) + k_2 \cos(\chi)}{2 z} \Bigg\{1 + \frac{\lambda}{\pi^2 \left(k_1^2 + k_2^2\right)}
    \frac{1}{2 \sin(\psi_0 + \chi)} \nonumber \\
  &\quad \Bigg[\frac{\sin^2(\psi_0 + \chi)}{4 \cos^3(\psi_0 + \chi)} \left(\pi-2 \psi_0 - 2 \chi -  \sin(2 \psi_0 + 2 \chi)\right) \nonumber\\
  &\quad - \cos(\chi) \sin(\psi_0) \cos^4(\psi_0) - \sin(\chi) \cos(\psi_0) \sin^4(\psi_0) \nonumber \Bigg] \nonumber \\
  & \quad + \mathcal{O}\left(\frac{\lambda^2}{\pi^4 (k_1^2 + k_2^2)^2}\right)
    \Bigg\},
\end{align}
\begin{align}
  V^{(\mathrm{II})}
  &= -\frac{n \sin(\chi)}{2 \sqrt{2} z} \Bigg\{1 + \frac{\lambda}{\pi^2 n^2} \left[\frac{\sin(\chi)}{4 \cos^3(\chi)} \left(\pi - 2 \chi - \sin(2 \chi)\right) - \frac{1}{4}\right] \nonumber \\
  &\qquad \qquad \qquad + \mathcal{O}\left(\frac{\lambda^2}{\pi^4 n^4}\right) \Bigg\}.
\end{align}
We thus find perfect agreement with the field theory results to two leading orders in the double scaling limit. Notice also that when $\psi_0 \rightarrow 0$ (namely $k_2/k_1 \rightarrow0$), the expansion for the action in~\eqref{actionI} reduces to the result for the Wilson line in the D3-D5 case~\cite{Nagasaki:2011ue,deLeeuw:2016vgp}.  For the set-up II the correction 
to the potential looks similar to the one of the D3-D5 brane case up to a replacement of $n$ by $\sqrt{2}k$. This is a peculiarity of the
one-loop approximation where only the first term in the expansion in eqn.~(\ref{caseII}) contributes, and it will not remain true at 
higher loop orders.  
Finally, we mention that for the set-up I there is no particular point of symmetry where the potential vanishes.
This is due to the fact that for set-up I all scalar fields get vevs, and it is not possible to choose a direction on the
sphere which is unaffected by these.

\section{Conclusion and Outlook}
\label{sec:conclusion}
Our investigation of Wilson lines provides an example that the AdS/dCFT dictionary for non-local observables remains valid upon breaking of both supersymmetry and (boundary) integrability.  In addition, it  serves as an important  consistency check of the perturbative framework that was set up  in references~\cite{Grau:2018keb,Gimenez-Grau:2019fld} for the dCFTs involved. We stress that having a perturbative framework for these defect CFTs is indispensable as these theories, 
due to the lack of supersymmetry,  are not amenable to methods such as localization. For other defect versions of ${\cal N}=4$ SYM, conserving part of the supersymmetries, such as
the D3-D5 probe brane model, important progress on the use of localization has recently been made in~\cite{Wang:2020seq}.

With the perturbative framework and the AdS/dCFT dictionary  in place, possibilities for further scrutiny of the present defect CFTs open up.  F.inst.\ one can scan the parameter spaces of the models for the presence of Gross-Ooguri like
phase transitions~\cite{Gross:1998gk}  as it was done for the supersymmetric D3-D5 probe brane set-up in~\cite{Aguilera-Damia:2016bqv,Bonansea:2019rxh,Bonansea:2020}.
It would likewise be interesting to study the transport properties of the various defect CFTs, supersymmetric or not,
by calculating correlation functions of the stress energy tensor across the defect or other related quantities.

\vspace{0.2cm}

\section*{Acknowledgments}
S.B.\ was supported by ``Fondazione Angelo Della Riccia" and the University of Florence.
C.K.\ and M.V.\ were supported by DFF-FNU through grant number DFF-4002-00037. 
\vspace*{0.5cm}



\bibliographystyle{elsarticle-num}
\bibliography{references}




\end{document}